\documentstyle[12pt,fullpage]{article}
\begin{document}
\begin{titlepage}
\begin{center}
\setcounter{footnote}0
\bigskip
\bigskip
\bigskip
\bigskip
{\bf The statistical dynamics of thin vortex line.}\\
\bigskip
\bigskip
Dmitry Podolsky\footnote{E-mail address: podolsky@itp.ac.ru}\\
\bigskip
{\em L.D.Landau Institute for
Theoretical Physics
\\ 2, Kosygin st., 117334, Moscow, Russia}\\
\bigskip
{\em 24 June 2001}\\
\end{center}
\begin{abstract}
We discuss the statistical properties of a single vortex line in a perfect fluid.
The partition function is calculated up to the end in the thin vortex
approximation. It turns out that corresponding theory is renormalizable, and the
renormalization law for the core size of the vortex is found. 
The direction of renormalization group flow makes the thin vortex approximation
to be valid for the interest cases and this result does not depend on
the choice of infrared regularization.
The expressions for some gauge-invariant
correlators are obtained to demonstrate the developed formalism at work.
\end{abstract}
\end{titlepage}

\newpage

\begin{center}
{\bf \it To Lilia: Could I find a black pearl in the road dust?}
\end{center}
\bigskip

\section{Introduction.}

Is it relevant to say that the phenomenon of turbulence is possible in a
perfect fluid?
How much is important for the turbulence to be realized
that real liquid has viscosity? We do not have the answers based on the theoretical analysis
yet. 

One of the intrinsic properties of a perfect fluid is the existance of
an infinite number of integrals of motion
providing
the lines of vorticity to be frozen in a liquid (see for example
\cite{ZK97},
 \cite{M98} and references therein).
The "freezing" means the possibility to mark every vortex line in such a way
that these lines are deforming with time, but the labeling remains always
the same.

Taking a volume element of a liquid, filled by vortex lines densely
and attempting to move one of its parts with respect to another, one
can feel the resistance to this motion because of the quasi-Coulomb
interaction between vortices. That is the turbulent viscosity.
It seems to be that
the fundamental viscosity is not important provided there is a lot
of vortices in the volume element and they are packaged densely ---
the main contribution to turbulent viscosity is due to the interaction
between vortices.
Small fundamental viscosity should result in the violation of "freezing"
condition: the conservation of integrals freezing lines of vorticity
will be destroyed with time.
But if the time of mixing is much less than the time on which these
integrals of motion are changing essentially, 
then the liquid is effectively ideal \cite{Ber}.

Nevertheless, the other effect depending strongly on the value of fundamental viscosity
becomes important in this case.
That is the formation of finite - time singularities in the solutions of equations of motion for
smooth initial conditions. The formation of such singularities really
may take place for the Euler equation (see for example
\cite{RP2001PRE}). As for the Navier-Stokes equation,  the situation is not clear yet.
We shall not take into consideration the question
about finite-time singularities in the present work.

The following program seems to be natural when the theoretical analysis of the 
situation is absent. First, we have develop the theory of the statistical hydrodynamics
in the case of a perfect fluid and to calculate the measurable quantities such as simple correlators.
Then, we should compare the theoretical predictions and experimental results.
From this point of view 
the problem of
statistical dynamics of a single vortex line in a
perfect fluid seems to be relevant. Two ways to its solution are known: the
approach developed in Refs. \cite{C1} - \cite{C5}
consists in postulating  Gibbs-like statistics for a vortex, and the other
approach is realised in the Ref. \cite{Ber}, where the probability of a
closed vortex line carrying the
vorticity $\Omega$ to be within a small tube surrounding some contour
$\gamma$ is calculated.

It was too hard to advance further than obtaining the results of qualitative nature \cite{C4}
within the framework of the first approach. Effectively the difficulties are due to the
necessity to calculate the generating functional for nonlocal quantum field theory.
In the present work we have calculated the generating functional completely in the thin
vortex approximation. It has been found that corresponding effective theory is renormalizable,
and the direction of renormalization group flow makes the thin vortex approximation to be valid.
The expressions for correlators connecting both
approaches for the solution of the problem have also been obtained.

\section{Calculation of the partition function.}

The Hamiltonian of a single vortex in perfect fluid has the following form
\begin{equation}
\label{Hcoul}
H = m_0 \int \int
\frac{\left(\vec{R}_{\xi_1} ' \cdot \vec{R}_{\xi_2}' \right)}
{ | \vec {R} \left(\xi_1 \right) - \vec{R} \left(\xi_2 \right)|}
d\xi_1 d\xi_2,
\end{equation}
here $ \xi $ is a parameter along the vortex line, components of $ \vec {R} \left (\xi \right) $ are
coordinates of the vortex
in the space, and $m_0 $ is  some constant with dimension of energy per unit length.

Keeping in mind the problem of statistical dynamics of a vortex and following the Refs. \cite{C1} - \cite{C5},
 we can write the partition function
\begin{equation}
\label{Zint}
Z = \int D ^ {*} \vec{R} \exp \left(-\beta H \right)
\end{equation}
Here the "star" in the measure of functional integral means that it is necessary to take into account
 an invariance of the Hamiltonian
(\ref{Hcoul}) with respect to the gauge transformation $ \tilde{\xi}  = f (\xi) $, where $f$ is an arbitrary function.
The theory defined by the Hamiltonian (\ref{Hcoul}) is strongly nonlocal, and the functional integral (\ref{Zint})
is essentially non-Gaussian, so it can be hardly calculated up to the end. It turns out that
it is possible to drastically simplify the problem using the well-known local induction approximation.
Its essence  consists in the following. Let us imagine that the "temperature" in the system $1/\beta $ is small enough
for the vortex to be deviated slightly from its equilibrium state.
Then the main contribution to the integral on $ \xi_2 $ from (\ref{Hcoul}) is given by the region, where
 $ \xi_2 \sim \xi_1 $, and the energy of the vortex appears to be proportional to its length with the logarithmic
 accuracy:
\begin{equation}
\label{LIcoul}
H \simeq m_0 \log \left( \frac{\Lambda}{l} \right) \int |R_\xi ' | d\xi,
\end{equation}
Here $l $ is the core size of the vortex, $ \Lambda $ is a typical size of the system in the case of an open vortex line
or the length of a vortex provided that it is closed.

It is convenient for us to expand the scope of applicability of the local induction approximation
changing the initial theory as follows (see the Ref. \cite{RP2001PRE}, but there is the opposite
sign of $\alpha$):
\begin{equation}
\label{LInc}
H = m_0 \int \int
\frac{\left(\vec {R} _ {\xi_1} ' \cdot \vec{R} _ {\xi_2} ' \right)}
{ | \vec{R} \left(\xi_1 \right) - \vec{R} \left(\xi_2 \right)| ^ {1 + \alpha}} d \xi_1 d \xi_2
\sim \frac{m_0}{\alpha} l^{-\alpha} \int |R_\xi ' | d\xi
\end{equation}
(the dimension of $m_0 $ varies correspondingly). This approximation is relevant because all the energy of a vortex
is concentrated in its core now, and the less the thickness of a core, the better it works.
Actually it is possible for such modified theory to permit strong fluctuations, if a vortex is thin enough.
The dynamics of $l $ also becomes unimportant in this case --- the waves moving along the vortex
and seeking to smooth a non-uniformity in the distribution of the thickness of the vortex line, have a speed,
which is proportional
 to $1/l $.

The answer for the partition function can be received from the lattice
representation \cite{P} of the theory without direct calculation of the functional integral. We have on the lattice
\begin {equation}
\label {Zlat}
Z = \sum_L c ^ {L/a} \exp \left (-m l ^ {-\alpha} L \right),
\end {equation}
where $m = \frac {m_0 \beta} {\alpha} $, $c ^ {L/a} $ is the number of paths with length $L $ on the lattice
 with the unit cell of characteristic size $a $, the constant
 $c $ depends on the concrete form of the lattice, the summation is performed over all configurations
 of the vortex line with length $L $.

As it is easy to see, this theory is remormalizable: the equation of a renormalization group is the independence condition
of the partition function from the parameters of the lattice
\begin {equation}
\label {Ren}
\frac {dZ} {da} = \frac {\partial Z} {\partial a} +
\frac {\partial Z} {\partial m} \frac {dm} {da} + \frac {\partial Z} {\partial l} \frac {dl} {da} = 0
\end {equation}

Generally speaking, $m $ can depend on the length of a vortex, as it takes place in (\ref {LIcoul}), and
it should be the absence of its renormalization.
Thus, the physical (observed) thickness of the vortex $l _ {ph} $ depends on the ultraviolet cut-off in
the following way:
\begin {equation}
\label {dren}
l = \left (l _ {ph} ^ {-\alpha} + \frac {\log c} {m} \frac {1} {a} \right)^{-1/\alpha}
\end {equation}

The calculation of the partition function (\ref {Zint}) performed without engaging the lattice
representation can be found in \cite {P}. The expression for the partition function of the open vortex
with ends fixed in the points $ \vec {R} _i $, $ \vec {R} _f $ is the following:
\begin{equation}
\label{PolZ}
Z = {\rm Const.} \int \frac{d^3 p}{p^2 + \mu} \exp
\left(i\vec {p} \left(\vec{R} _f - \vec{R} _i \right) \right),
\end {equation}
where
\begin {equation}
\label {mu}
\mu = \frac{1}{a} \left( m l ^ {-\alpha} - \frac{\rm Const.}{a} \right)
\end {equation}
and the value $a $ has the same sense, as for the lattice representation.
As it is easy to see, the renormalization law for the thickness of a vortex
(it can be found from the requirement of a finiteness of $ \mu $)
is similar to (\ref {dren}).

\section{Gauge-invariant correlators.}

Here it will be shown how this formalism is connected with approach developed in \cite{Ber}.
Let us calculate the probability for a vortex line to pass through a prescribed set of points
 $(\vec{R}_1, \ldots , \vec{R}_N)$. The physical meaning of this quantity is very close to
the meaning of probabilistic measure introduced in \cite{Ber} --- the probability for a vortex
line to be within some volume element $V$. This probability $F$ is also important because of its
universality: various physical quantities can be expressed in terms of $F$. We have \cite{P}
\begin{equation}
\label{F1}
F\left( \vec{R}_1 , \ldots , \vec{R}_N \right) = \left\langle \prod_j
\int_0^L d\xi_j \delta \left( \vec{R}\left( \xi_j \right) - \vec{R}_j \right) \right\rangle
\end{equation}

It is convenient to calculate this correlator using momentum representation:
\begin{equation}
\label{F2}
F\left( q_1 , \ldots , q_N \right) = \left\langle \prod_k \int_0^L d\xi_k \exp
\left( i \sum_j \vec{q}_j \vec{R} \left( \xi_j \right)     \right)  \right\rangle
\end{equation}

We omit here all the intermediate calculations (see Appendix for the details). The
final answer is (we suppose that $\sum_i \vec{q}_i = 0$ and 
$\sqrt{\mu} |\vec{R}_f - \vec{R}_i| \gg N$)
\begin{equation}
\label{F3}
F\left( q_1 , \ldots , q_N \right) \simeq {\rm Const.}
\int_0^1 \prod_{n=1}^N dx_n
\frac{\exp \left( i \sum_k \vec{q}_k \left( \vec{R}_f - \vec{R}_i \right) x_k \right)}{\left( \mu +
\frac{3}{4}D \right)^{3/4 + N/2}},
\end{equation}
where  $D = \sum_{j,k} \left( \vec{q}_j \cdot \vec{q}_k \right) D_c\left( x_j | x_k \right)$,
$D_c\left( x_j | x_k \right) = \sum_{n \ne 0} \frac{1}{2\pi n} \exp \left( i2\pi n \left( x_j - x_k  \right)  \right)$.

\section{Summary.}

We have discussed some statistical properties of a single vortex line under the proposal that its
probabilistic measure is Gibbs-like (\cite{C1} - \cite{C5}). The
partition function is calculated up to the end in the thin vortex approximation. It turned out that
the corresponding theory is renormalizable. The remormalization law for the core size of a vortex
line has been calculated, and the expressions
for some gauge-invariant correlators have been obtained to show the thin vortex approximation at work.

What is the meaning of these results?
First, it is easy to see from the representation of renormalization law (\ref{dren})
 that if the local induction approximation works well for the physical
thickness $l _ {ph} $, then it works even better for the "naked" thickness
$l $ from the Hamiltonian (\ref {LInc}). The following condition should be satisfied
 for our calculation of the number of configurations
in (\ref{Zlat}) to be correct:
\begin{equation}
\label{S1}
l_{ph}^{-\alpha} + \frac{\alpha \log c}{m_0 \beta a} \le a^{-\alpha}
\end{equation}
Then, it is possible to make more strong statement: the local
induction approximation works practically always for the "naked" thickness of the vortex $l $, even
when its using is not relevant for the physical thickness $l _ {ph} $. In fact, all the interest cases
of vortex structures should lay in the scope of applicability of this representation.

Second, the direction of renormalization group flow does not depend on the type of infrared
regularization. The only thing we need to know is that there is the "core" of our vortex --- a
transverse (with respect to the vortex line) scale where the energy of the vortex is localized,
then the thin vortex approximation can be used to calculate the partition function and correlators.

Thus, these results have the remarkable properties of universality.

\section*{Appendix.}
We will start from the following expression for the correlator $F\left( q_1 , \ldots , q_N \right)$:
\begin{equation}
\label{A1}
\begin{array}{l}
F\left( q_1 , \ldots , q_N \right) = \int_{|\vec{R}_f - \vec{R}_i|}^{+\infty}
dL \exp \left( -\frac{L}{2a\sqrt{\pi}} - md^{-\alpha}L + \frac{L}{a}  \right)   \\
\int \prod_{j=1}^N dl_j \int D\vec{R} \exp \left( -\int_0^L dl \frac{\left(R'\right)^2}{a} + i\sum_k \vec{q}_k \vec{R}\left( l_k \right)     \right)
\end{array}
\end{equation}
All the details concerning the way to deduce this expression can be found in \cite{P}.

First, we should perform the integration over $\vec{R}$. The integral is Gaussian, and following the usual
method, we have to solve the equation
\begin{equation}
\label{A2}
2\frac{1}{a}\vec{R}_{cl}'' = -i \sum_j \vec{q}_j \delta \left( l-l_j \right) =
i\sum_j \frac{a \vec{q}_j}{2} D \left( l|l_j \right)
\end{equation}
with the boundary conditions $\vec{R}_i = \vec{R}_{cl}(0) = i\sum_j \frac{\vec{q}_j A_j a}{2}$,
$\vec{R}_f = \vec{R}_{cl}(L) = i\sum_j \frac{\vec{q}_j B_j a}{2}$.
The equation for the Green function $D\left(l|l_j\right)$ is
\begin{equation}
\label{A3}
\begin{array}{rcl}
D'' \left(l|l_j\right) = -\delta \left(l - l_j\right), & D\left(0|l_j\right) = A_j, & D\left(L|l_j\right) = B_j
\end{array}
\end{equation}
Its solution is the following:
\begin{equation}
\label{A4}
D\left(l|l_j\right) = C_{1j} + C_{2j}l + \sum_{n\ne 0} \frac{1}{L\lambda_n}
\exp\left( i\sqrt{\lambda_n}\left(l - l_j\right)\right),
\end{equation}
where $\lambda_n = \frac{4\pi^2 n^2}{L}$.

The constants $C_{1j}$ and $C_{2j}$ are defined by the boundary conditions. The final answer is
\begin{equation}
\label{A5}
\begin{array}{l}
\vec{R}_{cl}(l) = \vec{R}_i - \frac{ia}{2}\sum_{j,n}\frac{\vec{q}_j}{L\lambda_n}
\exp\left(-i\sqrt{\lambda_n}l_j\right) + \frac{\vec{R}_f - \vec{R}_i}{L}l + \\
\frac{ia}{2}\sum_{j,n}\frac{\vec{q}_j}{L\lambda_n}\exp\left(-i\sqrt{\lambda_n}\left(l-l_j\right)\right)
\end{array}
\end{equation}

Now we have to calculate the classical action for this trajectory:
\begin{equation}
\label{A6}
\begin{array}{l}
S_{cl} = \int_0^L dl \frac{1}{a} \left( R_{cl}' \right)^2 - i\sum_k \left(\vec{q}_k \cdot \vec{R}_{cl}\left(l_k\right)\right)=\\
\frac{1}{a}\frac{\left(\vec{R}_f - \vec{R}_i\right)^2}{L} +
\frac{3a}{4}\sum_{j,k}\left(\vec{q}_j \cdot \vec{q}_k\right)
D_c \left(l_j|l_k\right) - i \sum_k \left( \vec{q}_k \cdot \left(\vec{R}_f - \vec{R}_i\right)\right)\frac{l_k}{L},
\end{array}
\end{equation}
where $D_c \left(l|l_k\right) = \sum_{n\ne 0}\frac{1}{L\lambda_n} \exp \left(i\sqrt{\lambda_n}\left(l-l_k\right)\right)$ ---
the Green function of the operator $\frac{d^2}{dl^2}$ on the circle.

Introducing the new variables $x_k = \frac{l_k}{L}$, we have
\begin{equation}
\label{A7}
\begin{array}{l}
F\left( q_1 , \ldots , q_N \right) = \int_{|\vec{R}_f - \vec{R}_i|}^{+\infty} dL
\prod_{n=1}^N \int_0^1 dx_n L^N \exp
\left(-\frac{L}{2a\sqrt{\pi}}-md^{-\alpha}L + \frac{L}{a} - \right. \\
\left. \frac{\left(\vec{R}_f - \vec{R}_i\right)^2}{aL} -
\frac{3}{4}aL \sum_{j,k} \left(\vec{q}_j \cdot \vec{q}_k\right) D_c \left(x_j|x_k\right) -
i\sum_k \left(\vec{q}_k \cdot \left(\vec{R}_f - \vec{R}_i\right)x_n\right)\right)
\end{array}
\end{equation}
The integral over $L$ can be calculated using the saddle point approximation (it work in fact always).
The answer is rather complicated in the general case, but it can be reduced to (\ref{F3}) in the limit
$\sqrt{\mu}|\vec{R}_f - \vec{R}_i| \gg N$.

\end{document}